# Safeguarding Decentralized Social Media: LLM Agents for Automating Community Rule Compliance

LUCIO LA CAVA, DIMES Dept., University of Calabria, Italy
ANDREA TAGARELLI, DIMES Dept., University of Calabria, Italy

Ensuring content compliance with community guidelines is crucial for maintaining healthy online social environments. However, traditional human-based compliance checking struggles with scaling due to the increasing volume of user-generated content and a limited number of moderators. Recent advancements in Natural Language Understanding demonstrated by Large Language Models unlock new opportunities for automated content compliance verification. This work evaluates six AI-agents built on Open-LLMs for automated rule compliance checking in Decentralized Social Networks, a challenging environment due to heterogeneous community scopes and rules. Analyzing over 50,000 posts from hundreds of Mastodon servers, we find that AI-agents effectively detect non-compliant content, grasp linguistic subtleties, and adapt to diverse community contexts. Most agents also show high inter-rater reliability and consistency in score justification and suggestions for compliance. Human-based evaluation with domain experts confirmed the agents' reliability and usefulness, rendering them promising tools for semi-automated or human-in-the-loop content moderation systems.

*Warning: This manuscript may contain sensitive content as it quotes harmful/hateful social media posts.*

## 1 Introduction

Social media platforms have become a primary channel of communication today, allowing anyone to easily create content and build communities around shared interests. However, this accessibility also comes with a cost: the spread of content that may be unsuitable for certain platforms or communities, including *hate speech*, *toxic content*, *harmful or violent content*, and *cyber-bullying*.

To address or mitigate this proliferation, most platforms declare community rules to establish the expected *compliance* that created content should meet in relation the community's members [41]. The purpose of community rules includes establishing guidelines for acceptable behavior, promoting safety and inclusivity, maintaining platform integrity, encouraging constructive engagement, and enforcing consequences for violations.

Compliance checking has traditionally been a manual task, with human volunteers and moderators chasing posts to assess their suitability. However, this is all but an even challenge: the number of moderators in a community is often way lower than needed, whereas the volume and virality of data are steadily increasing, thus limiting the compliance checking of just a sample of data in any amount of time, leaving most potential harmful content undetected. Besides, the workload associated with this volunteer effort has been found to cause significant psychological distress in moderators [43].

Platforms like Reddit attempted to alleviate such a burden by proposing the so-called *AutoMod*,[1] an automated *filter*-based mechanism to handle certain moderation tasks automatically. However, it has been perceived as "distant" to human expectations [21], thus making it not particularly suitable for complementing human moderators.

In the last couple of years, however, we have witnessed the emergence of very promising AI-based Natural Language Processing tools, particularly the Large Language Models (LLMs) [35, 53], which have the potential of paving the way for unprecedented capabilities in automatic rules compliance and content moderation. Indeed, such models have already been proven to be particularly effective in text annotation tasks, consistently outperforming humans, experts, and crowdworkers [14, 45]. Furthermore, the emergence of AI agents based on these LLMs [2, 49, 50] even more enabled human-like capabilities and adaptability within such systems, which can now better reason and collaborate.

---
[1]https://www.reddit.com/r/reddit.com/wiki/automoderator/

In this work, we aim to investigate the capabilities of LLM agents built on the most recently developed Large Language Models in the task of compliance rule checking and conformance alignment. Our goal with this work is to understand whether and to what extent LLM agents represent today a suitable choice for complementing and supporting humans in assessing content compliance with community rules, and we do this by means of the following research questions:

**RQ1** — *Are LLM agents able to perform compliance checking and suggest moderation strategies in communities with publicly available rules?*

**RQ2** — *Are there consistencies or differences in the behavior of LLM agents in delivering moderation strategies?*

**RQ3** — *How do humans (domain experts) perceive LLM agents' moderation strategies? Do they align well with human expectations?*

To answer these questions, we will focus on *Mastodon*,[2] the most well-known and widely used *Decentralized Social Media*, which represents a particularly challenging environment and benchmark for LLM agents, given the proliferation of heterogeneous community scopes and rules. Indeed, Mastodon allows anyone to create new servers tailored to certain topics (akin to Reddit subreddits), while the server administrators can declare a set of rules to be followed to stay compliant with the community's scope. Our choice of Mastodon is also supported by the public listing of rules, and the availability of *open* and *free-to-use* APIs (contrasting with most social platforms to date) that continue allowing researchers to obtain and analyze social data. Furthermore, this work was also motivated by the study from Hassan et al. [1], who explored the issue of *decentralized moderation*, unveiling a growing burden on server administrators in keeping their community healthy due to the growing user base, further exacerbated by the recent *#TwitterMigration* movement [18, 19, 28] following the Musk's acquisition of Twitter.

**Contributions.** Our contributions in this work are as follows: *(i)* We propose the first automatic compliance checking of social media content with community rules in the challenging context of Decentralized Social Media, thus differentiating from existing works typically focused on Reddit; *(ii)* in doing so, we propose the first large-scale evaluation of the compliance checking capabilities of six Open-LLM agents, contrasting with existing works typically focused on closed and API-based models; and *(iii)* through quantitative and human-based qualitative evaluations, we unveil that LLM-moderators are capable of detecting non-compliant content while adapting to different community context and rule set, thus being a suitable choice for novel AI-powered content moderation systems.

## 2 Related Work

### 2.1 Social Media Moderation

Keeping online social spaces safe has always been particularly challenging and intriguing for researchers from countless fields, aiming at fighting the proliferation of harmful content, and fostering a pleasant user experience [3, 16, 26].

Reddit has become the de-facto platform for community moderation and norm violation, thanks to the availability of public community rules and the strong effort in moderation carried out by volunteers. In this context, Fiesler et al. [12] explored the Reddit's rule ecosystem, Chandrasekharan et al. [9] shaped norm violation on Reddit using large-scale moderation insights, and Reddy et al. [41] characterized rules evolution in Reddit communities. Similarly, various studies assessed the effectiveness of community moderation on Reddit [47], further validating its effects [8, 20].

---
[2]https://joinmastodon.org/

Twitter/X emerged as the second most widely studied platform in this context; indeed, despite not having explicit community rules, it was characterized by a notable content-moderation effort. For instance, Zannettou [52] analyzed the effects of soft moderation, whereas Paudel et al. [37] proposed enhancements to the latter. Other works focused on the dynamics of account moderation [38] and the effects of Twitter moderation [22].

The development of LLMs paved the way for new approaches in content moderation, with the emergence of studies combining LLMs and content moderation. Among these, Franco et al. [13] were among the first to investigate the integration of LLMs in content moderation systems, focusing on the use of ChatGPT on Reddit. Similarly, Kolla et al. [25] studied the capabilities of GPT-3.5 in assisting Reddit content moderation. Recently, Kumar et al. [27] unveiled the high precision of GPT-3.5 in moderating Reddit subcommunities, also showing that (five) LLMs can better detect toxicity than humans.

It should be noted that, on the one hand, research was mostly focused on Reddit and Twitter, overlooking emerging platforms having a potential for content moderation, such as Decentralized Online Social Media, which are still to be properly investigated. On the other hand, automatic content moderation and rule compliance still remain at a preliminary stage and confined to Reddit only. Furthermore, despite the recent advancements in Natural Language Understanding, such initial efforts only considered GPT-3.5, thus leaving the plethora of Open LLMs unexplored, along with their positive implications on user privacy, due to the lack of black-box APIs.

## 2.2 Decentralized Social Media

Decentralized social media attracted the attention of researchers from various fields due to the novelty in their paradigm [15], and the implications decentralization might have on user-platform and user-user interactions [4, 10].

Mastodon stands as the most widely studied decentralized platform to date [6, 7, 40, 46, 54–56], due to similarities with micro-blogging platforms akin to Twitter/X, and the potential for a novel social paradigm, where horizontal growth between servers is preferred to the traditional vertical one, and incentivized to favor content diversification and community autonomy [56]. To date, the vast majority of existing studies involving decentralized social media and Mastodon exploit social networking aspects, delving into interactions among users or servers. As for the former, Zignani et al. [54, 55] were among the first to characterize Mastodon user interactions, aiming at finding similarities and differences with Twitter/X. They found that Mastodon exhibits a more balanced followers-followee distribution, with a more contained fraction of social bots, and the emergence of unusual disassortative traits. La Cava et al. [30] unveiled that decentralization impacts the development of user relations on Mastodon, leading to the emergence of strategic user roles and unique information consumption-production phenomena [31]. On the other hand, La Cava et al. [29] delved into the unprecedented relational front among servers of a single platform determined by Mastodon, unveiling and characterizing its structural footprint, and shaping polarization phenomena [32].

The establishment of decentralized social media platforms as a valuable alternative to their centralized counterparts, and the subsequent growth in user-base, brought *decentralized content moderation* under the spotlight. Hassan et al. [17] investigated the impact of decentralized content moderation on users in Pleroma, another decentralized social platform. Zia et al. [5] analyzed toxic content in Pleroma, also proposing a detection framework. Similarly, Nicholson et al. [36] characterized the rules from the most relevant Mastodon servers, comparing them to those of Reddit.

It should be noted that, despite these efforts in characterizing decentralized online social media, and the emerging interest in studying content moderation under such an unprecedented social scenario, the exploration of the potential for automatic moderation and its implications remained so far preliminary [25] in both scope and extent, hence prompting us to design a framework for LLM agent-based content moderation, as presented in the following sections.

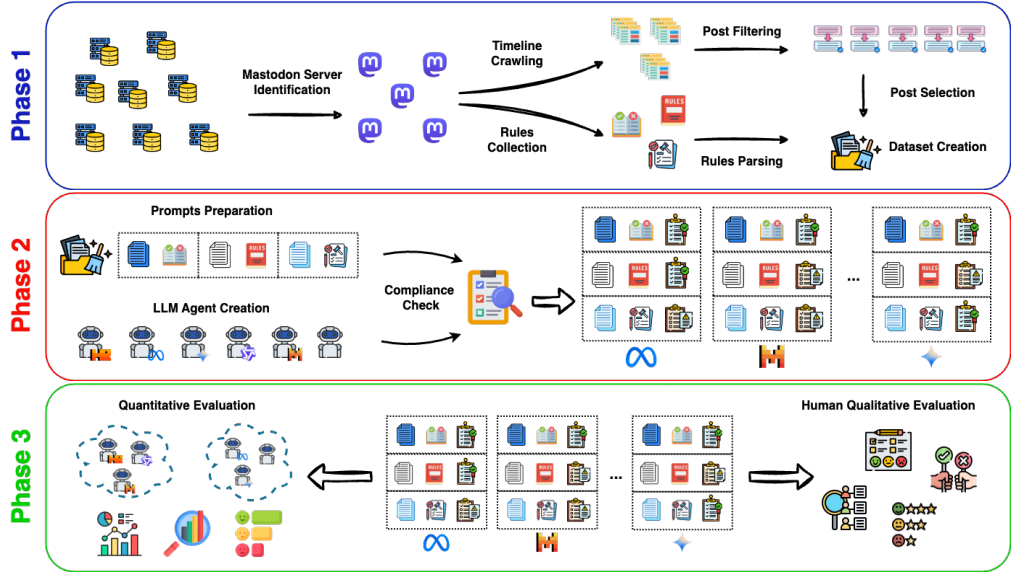

Fig. 1. Illustration of the main workflow of our proposed framework for automatic compliance checking.

## 3 Methodology

### 3.1 Overview

Our goal with this work is to assess whether and to what extent LLM agents are suitable for checking compliance with the rules publicly shared within an online community and, consequently, for recommending moderation strategies useful for that community's members. For the remainder of this paper, we will refer to the LLM agents assigned to the above task as *LLM-moderators*. Our designed methodology can be equally applicable to different social media platforms, however in this work we will demonstrate it on Mastodon as a representative of decentralized online social networks, which represent an important benchmark nowadays due to their high variability in community scope and rules, thus being particularly challenging for LLMs.

Figure 1 illustrates the workflow of our methodology, which consists of three main phases. The first phase is aimed at identifying official Mastodon instances between the Decentralized Social Media servers, collecting their posts and community rules, and preparing the dataset for the assessment (cf. Sections 3.2 and 3.3). The subsequent phase involves crafting a suitable prompting strategy for querying our LLM-moderators (Section 3.4) and deploying them reliably and efficiently (Section 3.5). Finally, in the third phase we evaluate the agents' performance through both quantitative and human-based qualitative assessments (Section 3.6).

Table 1. Language distribution across Mastodon instances as detected by *langdetect* and *FastText*.

| Language   | en     | de     | fr    | ja    | es    | zh-cn | it    | others | unknown |
|------------|--------|--------|-------|-------|-------|-------|-------|--------|---------|
| Count      | 678    | 140    | 56    | 49    | 31    | 16    | 13    | 98     | 219     |
| Percentage | 52.15% | 10.77% | 4.31% | 3.77% | 2.38% | 1.23% | 1.00% | 7.54%  | 16.85%  |

## 3.2 Data Collection

**Identifying Mastodon Instances.** We used the well-known *instances.social* platform,[3] the de-facto tracker for Mastodon instances, to retrieve a seed list of online instances in mid-2024. We filtered out instances with just one active user registered (in addition to the owner), to avoid the presence of noisy/test instances in our study. We hence obtained an initial raw list of 4378 potential Mastodon instances. However, it should be noted that *(i)* instances.social can mistakenly list some non-Mastodon servers, and *(ii)* the open-source nature of Mastodon allows for developing unofficial forks of the software. To avoid inconsistencies and ensure high reliability during our crawling processes, we decided to consider only the servers using the *official* Mastodon software. To this aim, we resorted to the `/api/v2/instance` endpoint in the official Mastodon APIs to gather server information from all candidate instances, and we filtered out all servers not having the official software[4] in the `source_url` field of the metadata. This process yielded a set of 3594 *verified* and *online* Mastodon instances.

**Crawling and Filtering Posts.** We started crawling posts from our filtered seed-list of instances using the official Mastodon `/api/v1/timelines/public` endpoint. In particular, from each instance, we fetched the latest 4000 posts, if available, and we required these posts to be *local*, i.e., generated by users from that instance, not as reposts of contents generated in other servers.

We emphasize that during this crawling step, we enforced a series of measures to preserve privacy, which include *(i)* discarding any user metadata while getting timelines, *(ii)* removing any *Personally Identifiable Information* (PII) from posts directly during crawling, and *(iii)* discarding any multimedia and non-textual content. Besides, we also enforced a series of strategies aimed at limiting servers' overload, e.g., limiting the number of requests per time slot.

At the end of this process, we obtained the requested timelines from 1120 out of 3594 instances, totaling 3,447,578 posts. This difference in the number of instances providing us with data was expected, as not all instances contain public posts, and some can also run earlier or outdated software versions, not compatible with the latest APIs we used.

As most language models are limited to English data during training (cf. Section 5), we decided to consider only English posts during our experiments. To this aim, we applied a double filtering process, involving both instances and individual posts. As for the former, we first considered the `description` field from the metadata extracted through the `/api/v2/instance` endpoint and combined it with the (optional) `extended_description` field obtained via the `/api/v1/instance/extended_description` endpoint. These fields contain all the information declared by instances' administrators to describe the scope, aim, and community of their servers. We hence exploited these details by using a combination of the `langdetect`[5] and `FastText` (with the `lid.176` model)[6] frameworks to detect the instances' languages. Table 1 summarizes the language distribution for the instances that properly answered our data request, with more than half of the instances being English ones.

---

[3]https://instances.social/
[4]https://github.com/mastodon/mastodon
[5]https://pypi.org/project/langdetect/
[6]https://fasttext.cc/

> **Rules enforced by *hci.social***
>
> (1) Treat everyone with respect and consideration; under the umbrella of respect, we expect all participants to be mindful of their speech and behaviors.
> (2) Communicate openly and thoughtfully with others and be considerate of the multitude of views and opinions that may be different than your own.
> (3) Be respectful and mindful in your critique of ideas.
> (4) Respect others' identities in full, e.g., using their specified pronouns.
> (5) Respect others' right to engage or disengage in conversation.
> (6) Accept responsibility to take action, as bystanders and advocates, to call out and report misbehavior and to hold each other accountable to these rules.
> (7) Do not engage in harassment in any form, including comments that target other participants based on characteristics such as gender, gender identity and expression, sexual orientation, race, ethnicity, age, ability, status, physical appearance, body size, or religion.
> (8) Do not engage in unwelcome personal attention, particularly when one individual has authority over the other.
> (9) Do not engage in persistent, unwanted attempts to contact another group member, particularly when one individual has authority over the other.
> (10) Do not deliberately intimidate, stalk, or threaten violence.
> (11) Do not engage in sustained disruption of online discussions.
> (12) Do not advocate for, or encourage, any of the sanctioned behavior described above.
> (13) Do not use this server if you are younger than 18 years old.
> (14) Do not post content unlawful in the United States or the State of New Jersey.

Fig. 2. Example of community rules collected from *hci.social*.

Concerning posts, we first leveraged the `language` field associated with each post's metadata as a preliminary filter for keeping only English ones. Then, we further filtered out the remaining non-English posts by using our combination of frameworks as in the case of instances' descriptions.

We iteratively applied our filtering method, by first excluding non-English instances, and then removing non-English posts from English instances. This process led to 1,510,816 English posts.

**Gathering Servers' Rules.** To keep the environment stick to the scope and avoid the proliferation of unwanted content, Mastodon administrators typically enforce rules and policies for their servers. These are publicly available, and users registering to a given instance are required to follow them.

We acquired the sets of rules enforced by Mastodon instances using the official Mastodon `/api/v1/instance/rules` endpoint. Except for a negligible fraction of instances without rules (∼ 7%), the average number of rules declared by our considered Mastodon instances is 7.04 ± 3.33, with a maximum of 26 rules. Besides, the mean number of words used in servers' rules averaged across all instances is 15.04 ± 8.18, with an instance peaking up to a mean of 41.5 words per rule.

We report the rules enforced by the *hci.social* community in Figure 2, whereas interested readers can refer to the Appendix A for additional insights.

Interestingly, rules can significantly vary in scope across servers; for instance, users can find servers strictly prohibiting Not Safe For Work (NSFW) content and others allowing for nudity, as well as servers limiting certain extremist political views along with ones more incline to free speech. Likewise, server rules may vary in detail levels, with certain servers enforcing very detailed (or even verbose) rules, and others limiting their set of rules to the bare minimum. It is worth noting that such diversity in rules declaration will represent a notable challenge for LLM agents, as they have to be flexible enough depending on the community scope to moderate.

Nonetheless, most instances exhibit some interesting common patterns: "content" is the most locally-frequent word (i.e., the most widely used word given each rule-set) for the 21% of the considered instances, followed by "violent" in the 9.3% of the instances. Notably, "content" also emerges as the most globally-frequent word (i.e., the word occurrences considering all the rules at once), followed by "harassment', "must", and "violent". Both unveiled patterns suggesting that most server administrators are particularly keen on keeping their communities safe from violent or harmful content, and to this aim, they typically declare rules that "must" be respected.

We finally also conducted the *Part-of-Speech* (POS) *tagging* of the rules to further characterize how they are articulated, finding that *singular or mass nouns* are the most widely used in rules declaration (e.g., "language", "content", "community", "violent", "respect") under both the same local and global perspectives used for the previous case.

### 3.3 Post Selection

To capture a diverse range of content interactions for a comprehensive yet computationally-efficient analysis, we focused on posts with the most level of engagement in addition to posts with the least level of engagement. The former represent viral content with a significant potential impact on the community, warranting close examination, whereas the latter are equally important, as they include content that may seem "marginal" and "escape" moderation, yet eventually spread over time.

Notably, Mastodon posts retrieved via APIs are accompanied by engagement metadata, including the received number of replies, reblogs (or reposts), and favorites (or likes). Figure 3 (left) reports the distribution of such engagement types for the posts we considered in this study. Apart from the favorites, the other engagement types are generally low, despite presenting some remarkable peaks. This is reasonably due to the presence of smaller instances or instances with lower engagement levels.

To better understand the engagement of Mastodon users, we measured it based on the following key aspects that characterize the user experience in this platform:

- *Replies* to posts only appear in the timeline and not directly on the profile of the replying user;
- *Reblogs* appear on both timeline and user profiles, thus representing the strongest form of engagement in terms of visibility;
- *Favorites* are mostly intended as a "save for later" functionality, thus remaining hidden and not publicly visible akin to platforms like X or Linkedin.

In light of the above considerations, we devised an aggregated *engagement score* ($E$) that linearly combines the three above aspects of engagement associated to each post: $E = N_{rep} + 2N_{reb} + 1/2N_{fav}$, where $N_{rep}$ represents the number of replies, $N_{reb}$ the number of reposts/reblogs, and $N_{fav}$ the number of favorites a post has received. We thus assigned each post in our dataset with the corresponding engagement score.

Figure 6 (right) illustrates the distribution of the aggregated scores across the data, whereby most posts exhibit mid-to-low engagement, with some outliers showing exceptionally high values. To further refine our data, we first filtered out posts with no engagement at all, eliminating half a million posts. Indeed, while these posts may still warrant moderation, they introduce noise into the data, and currently have no reach. Additionally, we excluded instances with fewer than 100 posts, to avoid focusing on cases where the distinction between high and low engagement becomes meaningless due to insufficient data.

We also applied a filter based on post length, excluding excessively short posts (e.g., single-word posts like hashtags) and very long ones, typically consisting of copy-pasted stories or articles, rather than original content. To this aim, we only retained posts whose length fall in the $[10th, 90th]$ percentile range of character count. This resulted in 795,673 posts meeting our criteria. Among these posts, we finally selected the top-50 and bottom-50 posts by engagement from each instance to ensure a manageable sample for in-depth analysis, as shown in the following Sections. Our final dataset consisted of 50,800 posts.

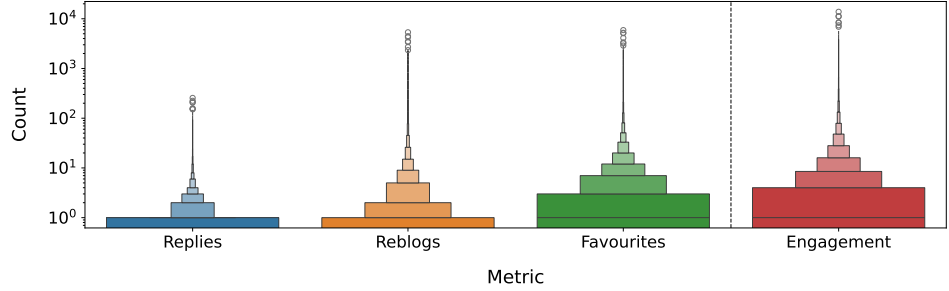

Fig. 3. (On the left) Distribution of different types of engagement in the raw Mastodon data. (On the right) Distribution of the aggregated-weighted engagement score associated with Mastodon data.

### 3.4 Prompting Strategy

To guide our LM agent moderators toward effective compliance checking capabilities, we crafted a prompting scheme designed to equip agents with all the necessary information for thorough moderation. These include establishing a persistent *system prompt message* aimed at conditioning their outputs to make them consistent throughout all their life cycles and generations.

More importantly, we presented each model with a *moderation instruction prompt* outlining the main guidelines for moderating content for the posts of any given instance, which are listed as follows:

- Details about the task to be performed,
- The list of the content policies enforced by a given Mastodon instance,
- The textual content of the post to be moderated,
- Instructions on the expected output format to generate, which include a *Compliance score* (to compute based on a Likert-scale), a *Justification* for the given score, and potential suggestions for *improving* the compliance.

Additionally, to reflect the availability of *content warnings* on Mastodon, we informed the agents whether these have been used, also including any *spoiler-text* used to hide the content. Figure 4 illustrates our prompting scheme.

### 3.5 LLM Moderators

**Models.** To create our set of LLM-moderators and assess their capabilities in moderating social media content, we resorted to a representative body of *open source* models, with varying sizes and architectures, and all of their implementations publicly available on the *HuggingFace Model Hub* as of mid-2024. It should be emphasized that we used open yet relatively small models due to the following reasons: *(i)* these models can be deployed locally (e.g., on the same server hosting the Mastodon instance), hence strictly ensuring privacy by avoiding sending content throughout the Web; *(ii)* these models come with better cost-effectiveness than commercially-licensed ones (e.g., GPT models); and *(iii)* their "openness" guarantees higher transparency than commercially-licensed ones, which are typically accessible as black-boxes hidden behind pay-to-use APIs.

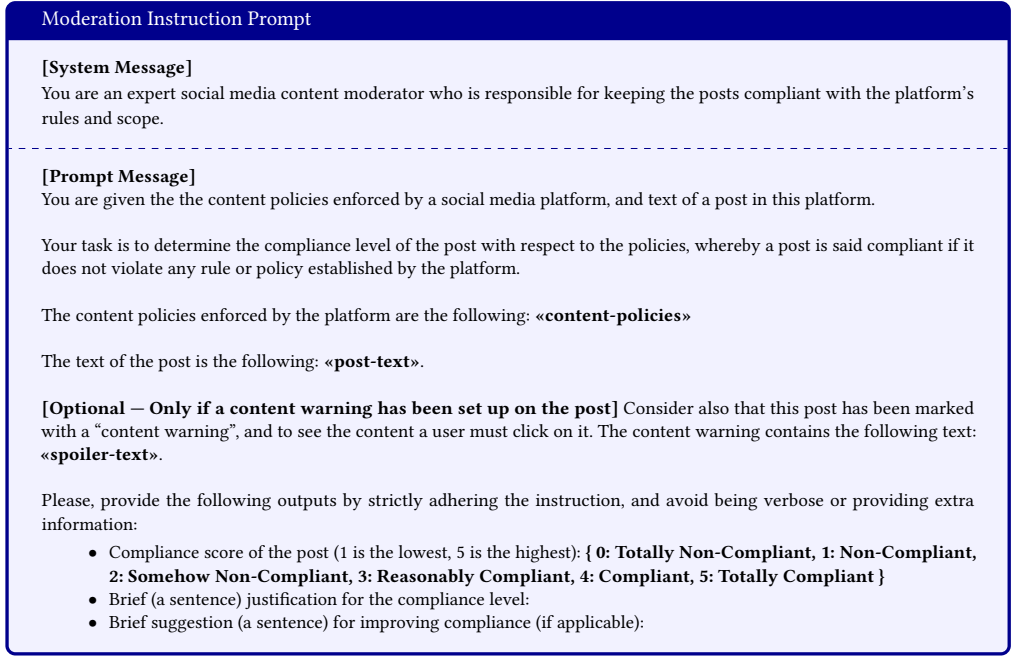

Fig. 4. Prompt containing the moderation instructions to be followed by the LLMs.

Table 2. Set of Large Language Models used in our study as *moderators*. Models are reported in decreasing order of size (number of parameters). For each model, we report the HuggingFace ID and the base architecture upon which it was developed.

| Model | ID | Params | Baseline |
|---|---|---|---|
| `Mistral-Nemo-Instruct-2407` | MistralNemo | 12.2B | Mistral |
| `gemma-2-9b-it` | Gemma2 | 9.24B | Gemma |
| `Meta-Llama-3.1-8B-Instruct` | Llama3 | 8B | Llama-3 |
| `Qwen2-7B-Instruct` | Qwen2 | 7.62B | Qwen |
| `Mistral-7B-Instruct-v0.3` | Mistral | 7B | Mistral |
| `Neural-chat-7b-v3-3` | NeuralChat | 7B | Mistral |

Table 2 reports the main characteristics of the LLMs selected in this study, namely *Mistral* [23] in its 7B and 12B variants, *Gemma* [44], *Llama3* [11] in its 8B variant, *Qwen2* [51] in its 7B variant, and Intel *NeuralChat*.

**Constrained Generation.** To ensure our moderators respect the output format we required through our prompting strategy, we used the *Guidance*, library[7] which allows *constrained generation* [34] (via the `select` method) to produce *structured* outputs strictly adhering to shape requirements; for example, this ensures that the moderator will produce a compliance score ranging within the Likert scale as required in the prompt. Similarly, for open-ended points, the Guidance framework allows us to keep the temperature value to the minimum allowed value for any particular model, to prevent models from hallucinating while commenting on the given scores. Additionally, we kept the *top_p* and *top_k* parameters to their default values of 50 and 1, respectively, to ensure reproducibility and enhance model stability — it is typically not recommended to tune both temperature and *top_p* to avoid descriptive effects on the generated outputs.

---
[7]https://github.com/guidance-ai/guidance

**Deployment.** The LLM-moderators were deployed locally through the Guidance framework and the Transformers back-end using an 8x NVIDIA A30 GPU server with 24 GB of RAM each, 764 GB of system RAM, a Double Intel Xeon Gold 6248R with a total of 96 cores, and Ubuntu Linux 20.04.6 LTS as operating system.

### 3.6 Assessment Criteria

Assessing the effectiveness of our LLM agents as social media moderators is challenging since Mastodon does not grant access to historical moderation events. Nonetheless, we overcome this through an in-depth evaluation framework at both quantitative and qualitative levels, which also includes *human-based evaluation* on the LLM agents' capabilities in keeping social media platforms safe and compliant with community rules. Next, we elaborate on the various steps of our evaluation methodology.

*3.6.1* **Quantitative Evaluation.** In our study, each LLM-moderator independently assigns compliance scores to posts. This hence makes it necessary to measure the level of agreement and the potential emergence of consensus among agents in moderating posts.

**Inter-Agent Agreement.** To measure consistency among agents, we first resorted to the *Fleiss's kappa coefficient* ($\kappa_F$) a well-established and robust statistical measure of inter-rater agreement between 3 or more raters. This measure is particularly suited for our compliance scores, which are categorical data on a *Likert* scale.

Given a set of $N$ LLM-moderators (i.e., raters) assigning posts with categorical scores, the Fleiss's kappa is measured as:

$$\kappa_F = \frac{\bar{P} - \bar{P}_e}{1 - \bar{P}_e}$$

where $\bar{P} = 1/N \sum_{i=1}^{N} P_i$ is the average proportion of agreement among raters, and each $P_i$ is computed as:

$$P_i = \frac{1}{N(N-1)} \sum_{j=1}^{h} \left( n_{ij}(n_{ij} - 1) \right)$$

where $n_{ij}$ is the number of LLM-moderators that assigned the $i$-th post to the $j$-th category (i.e., compliance score), with $j = 0..h$ (i.e., $h = 5$ in our Likert-scale). Also, $\bar{P}_e = \sum_{j=0}^{h} p_j^2$ is the average expected proportion of agreement by chance, with $p_j$ indicating the proportion of assignment to the $j$-th category across all considered posts.

We also resorted to the *Cohen's kappa coefficient* ($\kappa_C$) to measure pairwise inter-rater agreement. The rationale for using Cohen's kappa alongside Fleiss's kappa is that the former can reveal specific dynamics between pairs of LLM-moderators that may not be evident from group-level statistics. Cohen's kappa coefficient requires the raters to assess exactly the same items (i.e., posts), which lends itself particularly well to our setting. Given any two raters, we compute $\kappa_C$ as follows:

$$\kappa_C = \frac{p_o - p_e}{1 - p_e}$$

where $p_o$ is the observed proportion of agreement between the two raters (i.e., the fraction of posts whereby the two LLM-moderators gave the same compliance score), and $p_e$ is the expected proportion of agreement due by chance:

$$p_e = \sum_{j=0}^{h}(p_{j1} \times p_{j2})$$

with $p_{j1}$ and $p_{j2}$ indicating the marginal probabilities of rater-1, resp. rater-2, assigning a rating of $j$ (i.e., the proportion of posts rater-1, resp. rater-2, rated as $j$).

Both Fleiss's kappa and Cohen's kappa take values between -1 (perfect disagreement), to 1 (perfect agreement), with 0 indicating a level of agreement to be expected as driven by chance.

**Lexical Similarity.** As part of our prompting strategy, LLM-moderators are tasked with delivering open-ended justifications and suggestions for improvement related to the scores they assign to posts. To evaluate the agreement between agents in commenting on their outputs, we first measured *lexical similarity*, according to the overlap ($WO$) in the word-set used by the LLMs in their justifications, resp. improvement suggestions, associated with each post:

$$WO(t_1, t_2) = \frac{|t_1 \cap t_2|}{\min(|t_1|, |t_2|)}$$

where $t_1, t_2$ are the two LLM outputs which were pre-processed by performing lowercasing, punctuation removal, stop-word removal, and lemmatization. Clearly, the higher the overlap, the more similar the usage of specific words among two agents.

**Semantic Similarity.** Different LLMs might express similar justifications or suggestions while using different vocabularies, therefore we also measured the semantic similarity of the LLM-produced contents. To this purpose, we resorted to an approach based on *Sentence Transformers* [42], the de-facto standard for producing deeply-contextualized sentence embeddings for semantic search and similarity tasks. The semantic similarity for any two texts is then computed as the cosine similarity measure between their respective embeddings produced by the Sentence Transformer. In our experiments, we used `all-mpnet-base-v2`, which is widely known as one of the most effective Sentence Transformer models.[8]

**Assessing Biases.** We are also interested in evaluating the LLM-moderators' behavior in terms of a number of aspects that can affect the produced scores and justifications/recommendations. A list of such aspects and related questions they may arise is as follows:

- *Rules complexity*: is there a relation between the complexity of rule definitions in communities (e.g., verbosity) and their corresponding moderation scores?
- *Engagement*: how does the engagement level of a post relate to its compliance score?
- *Sensitive Content*: does admitted sensitive content impact on the moderation scores?
- *Score laziness*: is there a relation between the compliance score for a post and the length of justification/suggestion for improvements provided?
- *Community bias*: is there any bias toward certain community aspects (e.g., main topic) in assigning compliance scores?

To answer these questions, we carried out a correlation analysis whose results are discussed in Section 4.1 - Moderation biases.

---

[8]https://huggingface.co/sentence-transformers/all-mpnet-base-v2

> **Human-based Qualitative Evaluation Scheme**
>
> Community rules: **«community-rules»**
>
> Text of the post: **«post-text»**.
>
> Choose the most appropriate moderation model among the following:
> **«($score_1$, $justification_1$, $suggestion_1$), ($score_2$, $justification_2$, $suggestion_2$), ..., ($score_h$, $justification_h$, $suggestion_h$)»**.
>
> Based on your selected moderation model, please evaluate the following points:
> - Rate how well the given compliance score corresponds to your perception of the compliance of the post w.r.t. the rules: **«Likert[0-5]»**
> - Rate how well the justification text provided by the selected moderation model fits with the compliance score assigned by the selected moderation model: **«Likert[0-5]»**
> - Rate how much useful you evaluate the moderation approach of the selected model: **«Likert[0-5]»**
>
> Comments/Feedback on the strengths of the chosen moderation strategy: **«text here»**
> Comments/Feedback on the weaknesses of the chosen moderation strategy: **«text here»**

Fig. 5. Scheme for the human evaluation by our domain experts as questionnaire respondents.

*3.6.2* **Human-based Qualitative Evaluation.** Our final evaluation step involved assessing whether and to what extent humans perceive automatic moderation strategies enforced by our LLM agents as appropriate and close to their expectations. To achieve this, we designed a *survey for human evaluators*, as detailed below.

Given the time-consuming and expensive nature of manual annotation, along with the relatively large amount of posts (50,800) rated by our LLM-moderators, we opted to focus on a manageable sample of $M = 30$ posts, to be manually annotated by $N = 30$ human evaluators. This sample size balances resource constraints with the expected reliability of the annotations. It should be noted that the scores generated by our models (cf. Figure 6) are not uniformly distributed, with compliant posts being dominant. However, since moderation tends to be more challenging for non-compliant posts, we binned posts into 5 score ranges (cf. Appendix C), and sampled 6 posts from each bin uniformly at random. *Please note that, to avoid exposing annotators to extremely sensitive (e.g., disgusting/harmful/highly offensive) content, we manually filtered out certain posts from the ones in the lowest bins, replacing them using the same sampling strategy.*

The survey was administered using the *Qualtrics* platform.[9] Each human annotator was first presented with an introductory page explaining the scope, context, and instructions for taking the questionnaire (cf. Appendix C). Subsequently, for each sampled post, we filled in the template reported in Figure 5 with *(i)* the list of rules enforced by the server/community, *(ii)* the text of the original post (we recall to be PII-free), and *(iii)* the list of moderations suggested by our LLM agents — including scores, justifications, and suggestions for improvement. To prevent biases in evaluation, *this list was shuffled and anonymized, so that the human annotators were unaware of which LLM agent generated each strategy.* Annotators were then tasked to choose the most suitable moderation approach and express quantitative (on a Likert scale) and qualitative (using open-ended text fields) feedback. A screenshot of a sample page shown to users is reported in Appendix C.

This process allowed us to rank the performance of our LLM-moderators and gain insights into user preferences, which could be used in the foreseeable future to enhance and refine automatic moderation quality, e.g., by steering LLMs toward certain preferred strategies (cf. Section 5).

---
[9]https://www.qualtrics.com/

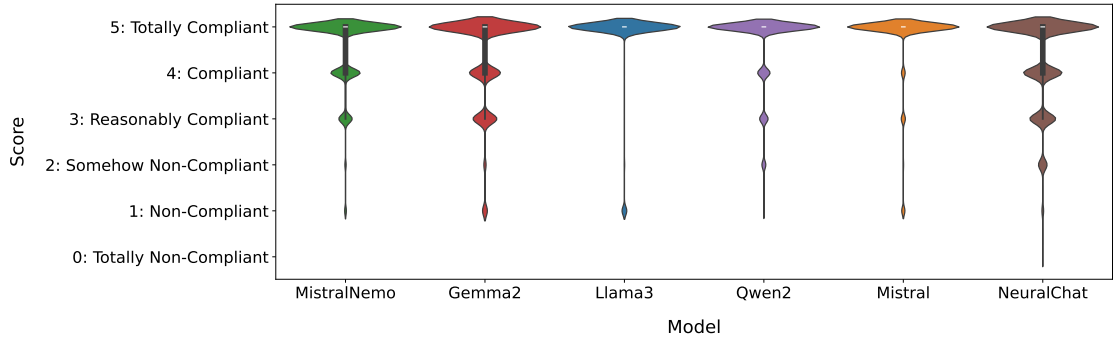

Fig. 6. Distribution of compliance scores assigned by the LLM-moderators to the posts considered in our study.

The human annotators were recruited among *domain experts* in Mastodon (including instance administrators) and experts in community moderation. We point out that we deliberately avoided using platforms like *MTurk*, as they have been found to be "contaminated" by widespread LLM usage [48], thus potentially undermining the significance and validity of the results. Furthermore, we prioritized high-quality feedback from domain experts to ensure high sensitivity w.r.t. the task, paramount for this first endeavor in automatically moderate decentralized social communities.

## 4 Results

In this section, we present an in-depth analysis of the results produced by our LLM moderators, including their evaluation by human reviewers.

### 4.1 Quantitative Results

**Compliance Score Assignment.** We started our analysis of the results by looking at the scores assigned by each LLM-moderator to the Mastodon posts, as shown in Figure 6.

It stands out that the most frequently assigned compliance score across all agents is "5: Totally Compliant". This is totally expected given the volume of data we collected per instance. Indeed, for smaller servers, this often involves reviewing posts from several months back, increasing the likelihood that non-compliant content has already been addressed by administrators or moderators. Further investigations on this point confirmed this hypothesis, revealing that posts receiving the lowest scores from our models tend to be more recent (cf. Figure 15 in Appendix B).

Beyond this general trend, we notice varying behavior in how the various models score the posts' compliance. For example, Llama3 appears to only assign full-compliance scores, with a very small number of non-compliant scores. Notably, NeuralChat is the only model having considered the "0: Totally Non-Compliant" option. Among the other models, MistralNemo, Gemma, and NeuralChat exhibit a wider range of assigned compliance scores.

This hence prompted us to quantitatively explore the extent to which the LLM-moderators agree in assigning compliance scores by leveraging the inter-agent agreement measures introduced in Section 3.6.1. We first obtained a Fleiss's $\kappa$ of 0.253, indicating a fair level of agreement among all considered raters [33]. This is a preliminary hint at a non-negligible consistency in assigning compliance scores exhibited by our selected models. However, some models may agree or disagree more than others.

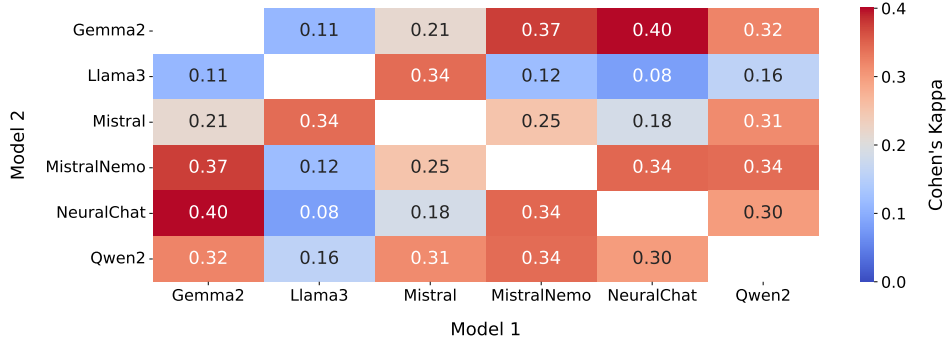

Fig. 7. Pairwise Cohen's $\kappa$ scores between the LLM-moderators.

Table 3. Average word overlap between the justifications (top) and suggestions (bottom) provided by the LLM-moderators.

|  |  | MistralNemo | Gemma2 | Llama3 | Qwen2 | Mistral | NeuralChat |
|---|---|---|---|---|---|---|---|
| Justification | MistralNemo | — | 0.456 ± 0.209 | 0.410 ± 0.190 | 0.451 ± 0.200 | 0.536 ± 0.220 | 0.476 ± 0.202 |
| | Gemma2 | 0.456 ± 0.209 | — | 0.393 ± 0.187 | 0.447 ± 0.195 | 0.570 ± 0.250 | 0.505 ± 0.194 |
| | Llama3 | 0.410 ± 0.190 | 0.393 ± 0.187 | — | 0.290 ± 0.138 | 0.450 ± 0.228 | 0.308 ± 0.146 |
| | Qwen2 | 0.451 ± 0.200 | 0.447 ± 0.195 | 0.290 ± 0.138 | — | 0.515 ± 0.223 | 0.399 ± 0.180 |
| | Mistral | 0.536 ± 0.220 | 0.570 ± 0.250 | 0.450 ± 0.228 | 0.515 ± 0.223 | — | 0.574 ± 0.254 |
| | NeuralChat | 0.476 ± 0.202 | 0.505 ± 0.194 | 0.308 ± 0.146 | 0.399 ± 0.180 | 0.574 ± 0.254 | — |
| Suggestions | MistralNemo | — | 0.737 ± 0.431 | 0.427 ± 0.484 | 0.165 ± 0.359 | 0.841 ± 0.358 | 0.292 ± 0.444 |
| | Gemma2 | 0.737 ± 0.431 | — | 0.226 ± 0.246 | 0.174 ± 0.332 | 0.612 ± 0.358 | 0.181 ± 0.225 |
| | Llama3 | 0.427 ± 0.484 | 0.226 ± 0.246 | — | 0.306 ± 0.267 | 0.451 ± 0.407 | 0.238 ± 0.252 |
| | Qwen2 | 0.165 ± 0.359 | 0.174 ± 0.332 | 0.306 ± 0.267 | — | 0.333 ± 0.362 | 0.454 ± 0.356 |
| | Mistral | 0.841 ± 0.358 | 0.612 ± 0.358 | 0.451 ± 0.407 | 0.333 ± 0.362 | — | 0.378 ± 0.425 |
| | NeuralChat | 0.292 ± 0.444 | 0.181 ± 0.225 | 0.238 ± 0.252 | 0.454 ± 0.356 | 0.378 ± 0.425 | — |

To investigate this further, we analyzed the agreement between specific pairs of raters to identify any notable patterns of stronger or weaker alignment beyond Fleiss's $\kappa$ value. Figure 7 reports the Cohen's $\kappa$ values computed for each pair of LLM-moderators to frame the degree of inter-rater agreement. Interestingly, most pairs exhibit *fair* to *moderate* strength of agreement [33], with MistralNemo, Gemma, and NeuralChat settling around $\kappa_C = 0.4$, thus confirming our earlier observations. It should also be noted that, although on lower values, the remaining pairs of moderators still exhibit a slight level of agreement [33]. Notably, Llama3 demonstrates the least agreement with the other models, due to its degeneracy in assigned scores.

Overall, our findings in inter-rater agreement suggest that *(i)* LLM-moderators share a similar understanding of the community rules to be enforced, leading to *(ii)* consistent capabilities of moderation.

**Score Justification and Suggestions for Improvement.** The moderate degree of agreement we spotted in assigning compliance scores among the LLMs prompted us to assess whether and to what extent such consistency also extends to the justifications for the given scores, as well as suggestions for improving compliance. In this regard, we analyzed the word overlap between the justifications/suggestions provided by each model (cf. Table 3), as well as the cosine similarity between their embedded representations (cf. Table 4).

Table 4. Average cosine similarity between the embeddings of the justifications (top) and suggestions (bottom) provided by the LLM-moderators.

|  |  | MistralNemo | Gemma2 | Llama3 | Qwen2 | Mistral | NeuralChat |
|---|---|---|---|---|---|---|---|
| Justification | MistralNemo | — | 0.670 ± 0.155 | 0.572 ± 0.141 | 0.659 ± 0.131 | 0.700 ± 0.156 | 0.668 ± 0.130 |
| | Gemma2 | 0.670 ± 0.155 | — | 0.555 ± 0.132 | 0.649 ± 0.122 | 0.700 ± 0.142 | 0.667 ± 0.128 |
| | Llama3 | 0.572 ± 0.141 | 0.555 ± 0.132 | — | 0.587 ± 0.137 | 0.591 ± 0.135 | 0.566 ± 0.133 |
| | Qwen2 | 0.659 ± 0.131 | 0.649 ± 0.122 | 0.587 ± 0.137 | — | 0.696 ± 0.121 | 0.685 ± 0.127 |
| | Mistral | 0.700 ± 0.156 | 0.700 ± 0.142 | 0.591 ± 0.135 | 0.696 ± 0.121 | — | 0.712 ± 0.127 |
| | NeuralChat | 0.668 ± 0.130 | 0.667 ± 0.128 | 0.566 ± 0.133 | 0.685 ± 0.127 | 0.712 ± 0.127 | — |
| Suggestions | MistralNemo | — | 0.482 ± 0.180 | 0.208 ± 0.128 | 0.336 ± 0.271 | 0.613 ± 0.349 | 0.265 ± 0.144 |
| | Gemma2 | 0.482 ± 0.180 | — | 0.248 ± 0.130 | 0.313 ± 0.155 | 0.423 ± 0.156 | 0.315 ± 0.136 |
| | Llama3 | 0.208 ± 0.128 | 0.248 ± 0.130 | — | 0.411 ± 0.159 | 0.329 ± 0.182 | 0.385 ± 0.164 |
| | Qwen2 | 0.336 ± 0.271 | 0.313 ± 0.155 | 0.411 ± 0.159 | — | 0.455 ± 0.262 | 0.494 ± 0.191 |
| | Mistral | 0.613 ± 0.349 | 0.423 ± 0.156 | 0.329 ± 0.182 | 0.455 ± 0.262 | — | 0.402 ± 0.254 |
| | NeuralChat | 0.265 ± 0.144 | 0.315 ± 0.136 | 0.385 ± 0.164 | 0.494 ± 0.191 | 0.402 ± 0.254 | — |

The moderate-to-low word overlap reported in Table 3 (top) suggests that the models use different jargon in justifying their assigned compliance score, as expected given the different underlying architectures (cf. Table 2) and vocabulary. Interestingly, Table 3 (bottom) shows a noisy overlap in terms of suggestions for improving compliance. Indeed, models were found to show significant overlap yet with high standard deviations. We attribute the high mean values to the zero-suggestion expressions, such as "N/A", "No need for improvement", "The post is already compliant", etc., which are frequent due to the large number of compliant posts. By contrast, the high deviation reveals that, when models generate more structured outputs, they tend to use distinct terms.

Concerning the semantic similarity between the justifications of the compliance scores, Table 4 (top) shows a lower bound of around 0.57 (i.e., Llama3 vs. MistralNemo). This suggests that, despite using different jargon as previously spotted, the meaning of the justification among models is rather aligned and consistent, even for models that exhibited a lower inter-rater agreement. More interestingly, the highest semantic similarity (around 0.7) is found for the models with the better inter-rater agreement (i.e., the Mistral family, Gemma2, and NeuralChat), further strengthening their consistency in moderation. Besides, Qwen2 also presents interesting similarities in justifications.

The semantic similarity decreases when considering the suggestions for improving compliance, as reported in Table 4 (bottom). Here, we spotted a more varied scenario, with agents yielding suggestions still having a fair relatedness (cosine similarity around 0.2), yet with a reduced strength. Even in this case, the Mistral family of models was the one with the higher consistency in suggestions (cosine similarity of 0.61), followed by Gemma2, NeuralChat, and Qwen2.

Our analysis of the lexical and semantic properties of the score justification and suggestion for improvement unveils that *(i)* while agents use different jargon for commenting on their assigned scores, they *(ii)* express similar semantic meanings, indicating a general agreement on the rationale behind their scores — further corroborating their inter-rater agreement. However, *(iii)* there is more variation in lexical and semantic nuances when it comes to suggesting improvements for enhanced compliance.

**Moderation biases.** A further evaluation We carried out is to understand whether and to what extent our LLMs perceive certain influence factors while checking compliance.

To begin with, we explored *whether the verbosity of community rules can affect moderation outcomes*, hypothesizing that more detailed rules could lead to stricter or more "severe" moderation. We have found no correlation between rule length and compliance scores, although with insufficient evidence (i.e., the no-correlation results might occur due to random chance), as indicated by Pearson's correlation $r = 0.054$, $p > 0.01$ and Spearman's rank correlation $r = -0.013$, $p > 0.01$.

Our second investigation involved discovering *whether post engagement is an influencing factor for compliance*, assuming that highly engaging content might be more compliant. Surprisingly, we found no correlation (Pearson's $r = -0.024$, $p < 0.01$; Spearman's $r = -0.020$, $p < 0.01$), which could reasonably explained by two factors: *(i)* non-compliant content can escape moderation, especially on larger instances with a higher volume of content, and become viral; and *(ii)* being compliant does not imply being viral.

We also analyzed the *potential emergence of "laziness"* in LLM agents when providing justifications and improvement suggestions alongside compliance scores. While we found no correlation between scores and justification length (Pearson's $r = 0.002$, $p > 0.01$; Spearman's $r = 0.031$, $p < 0.01$), suggesting that agents are not lazy in outlining the rationale behind a given score, we observed a slightly negative correlation (Pearson's $r = -0.163$, $p < 0.01$; Spearman's $r = -0.216$, $p < 0.01$) between compliance scores and the length of suggestions for improvement. This is particularly interesting, as indicates that agents tend to be more detailed when offering suggestions for improving compliance, especially for lower-scoring posts.

Despite excluding most potential biases in assigning compliance scores, we observed that *LLMs can be more "sensitive" when moderating NSFW content*. As noted in Section 3.4, agents were informed about the presence of sensitive content to be moderated, including the availability of a content warning and the corresponding spoiler text. However, we found a slightly negative correlation (Pearson's $r = -0.181$, $p < 0.01$; Spearman's $r = -0.206$, $p < 0.01$) between the presence of sensitive content and compliance scores. Although limited, this pattern can be ascribed to the built-in *safeguarding* mechanisms in LLMs, as well as to the possibility that even the spoiler texts over the content warnings may contain NSFW elements. Despite not being particularly evident, such an emerging sensitivity should be taken into account, as communities admitting NSFW content may face more severe automatic compliance checking outcomes. This sensitivity was further supported by our manual inspection that the instances with lower average compliance scores were typically communities centered around NSFW topics.

### 4.2 Exploratory Qualitative Analysis

Before conducting our human-based qualitative evaluation of the automatic compliance checking performed by LLM agents, we *manually* carried out an exploratory phase — focusing mainly on the *non-compliant* posts — aimed at identifying peculiarities and inconsistencies in the moderation strategies. From this analysis, we identified three main qualitative categories of moderation, which are described below as the *good*, the *bad*, and the *ugly*.

**The Good.** LLM agents were found particularly effective in automatic compliance checking, demonstrating impressive adaptability to diverse communities, and an ability to detect "latent" violations of community rules. They also demonstrated effectiveness in suggesting how to improve compliance, e.g., through rephrasing or removal of certain text portions.

Figure 8 illustrates an example of successful moderation by one of the LLMs. It can be noted that the model accurately identified the specific set of violated rules, precisely referring to them, and even pinpointing a specific nuance (i.e., referencing *"no hate speech"* within a broader rule). Furthermore, the model's suggestion for improving the compliance of the post w.r.t. the rules was clear and direct, underscoring the potential value of such an automatic compliance check. Notably, the model also took into account the "sentiment" expressed by the author (who however used inappropriate

4Safeguarding Decentralized Social Media: LLM Agents for Automating Community Rule Compliance 17---

**Qualitative Evaluation - Sample Post: "The Good"**

**Community Rules:**
*1. Don't Be A Ballox 2. No bigotry, racism, xenophobia, sexism, homophobia, transphobia, anti-Traveller speech or any hate speech. 3. No unsolicited DMs 4. We have Threads blocked, just so you know 5. We operate on a "3 strikes and you're out" basis.*

**Post Content:**
*I am exhausted hearing older folks saying the youth of today just aren't working hard enough. Go fucking fuck yourself.*

- - - - - - - - - - - - - - - - - - - - - - - - - - - - - - - - - - - -

**Moderation Strategy:**
**Score:** *1: Non-Compliant*
**Justification for the score:** *The post contains hate speech and violates policy 1 ("Don't Be A Ballox") and policy 2 ("No hate speech").*
**Suggestions for improvement:** *The post could be made more compliant by removing the offensive language and expressing the sentiment without using hate speech.*

---

Fig. 8. An example of "good" compliance checking by a sampled LLM-moderator.

language) and encouraged the author to convey her/his feelings in a more appropriate and compliant manner. This highlights the model's ability to engage with human emotions while maintaining rule compliance.

**The Bad.** However, despite these impressive capabilities, certain posts can still slip through the automatic compliance-checking process — similar to what can happen with human moderators. Figure 9 shows an example where a model failed to detect the use of *inappropriate* and *sexist* language in a post. Fortunately, all other models recognized the issue and proposed effective moderation strategies, in contrast to the one that missed it. In this regard, we point out that these cases are particularly rare and, when they occur, they are found to be limited to one model.

**The Ugly.** During our manual review of the moderation strategies generated by the agents, we encountered some "ugly" issues. First, LLMs could in principle hallucinate on the open-ended textual parts — while we were able to control and avoid this during score assignment thanks to constrained generation (cf. Section 3.4). Fortunately, this issue seems to be limited on Llama3, which generated some "noisy" moderation outputs, including content like portions of Python code or repeated phrases. Although these artifacts may introduce unnecessary clutter and lower the perceived quality of the output, it is important to note that these issues generally followed an initial informative and precise output. Moreover, the moderation scores themselves were found to be unaffected, remaining coherent and aligned with the rules.

Another interesting point to keep in mind when evaluating the quality of the moderation relates to the informativeness of the rule set declared by a server where the automatic compliance checker should be deployed. Indeed, we found that in the presence of poorly informative server rules (e.g., a single rule stating "be good"), LLMs may struggle in determining compliance, as the lack of clarity in the rules might lead to ambiguity. This can result in models hallucinating or suggesting generally accepted "good" behaviors that, while positive, might not align with the scope of the corresponding community, e.g., servers deliberately avoided declaring rules for fostering "free-speech".

### 4.3 Human Qualitative Assessment

Our final evaluation focused on assessing how humans, i.e., domain experts on Mastodon in our setting, perceive the effectiveness of the automatic compliance checks performed by our LLM-moderators. Below we discuss the key findings

> **Qualitative Evaluation - Sample Post: "The Bad"**
>
> **Community Rules:**
> *1. Sexually explicit or violent media must be marked as sensitive when posting 2. No racism, sexism, homophobia, transphobia, xenophobia, ableism, casteism, or any other form of bigotry. 3. No incitement of violence or promotion of violent ideologies 4. No harassment, dogpiling, doxxing or otherwise abusive behavior. 5. No illegal content 6. Do not share intentionally false or misleading information*
>
> **Post Content:**
> *Just for the record, I would 100% endorse a slut for President, because I base my choices on their values and the work they've done. You should sleep with as many consenting adults as you want, or none at all, why the fuck would I care?*
>
> - - - - - - - - - - - - - - - - - - - - - - - - - - - - - - - - - - - - -
>
> ***Bad* Moderation Strategy:**
> Score: *4: Compliant*
> Justification for the score: *The post does not violate any of the listed policies.*
> Suggestions for improvement: *N/A.*
>
> ***Proper* Moderation Strategy:**
> Score: *1: Non-Compliant*
> Justification for the score: *The post contains sexist language and promotes a derogatory term for women.*
> Suggestions for improvement: *Remove the derogatory term and refrain from using sexist language.*

Fig. 9. An example of "bad" compliance checking by a sampled LLM-moderator.

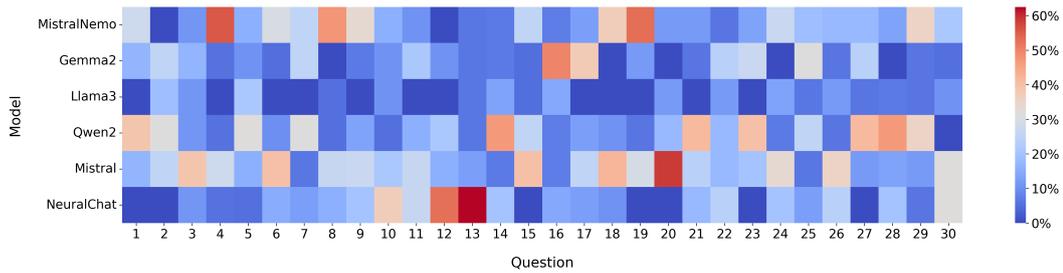

Fig. 10. Human preferences for each question (i.e., post to be annotated) based on the suggested moderation strategies. Warmer, resp. colder, colors indicate higher, resp. lower, percentage of preference of a model by the domain experts (questionnaire respondents).

from our administered survey, which gathered 900 annotations (i.e., 30 respondents providing feedback on 30 posts each).

**Human Preferences for LLMs' Moderation Strategies.** As described in Figure 5, we asked the questionnaire respondents to select their preferred moderation strategy from a shuffled and anonymized list of approaches generated by our LLMs. Overall, we found that the majority of questionnaire respondents favored the moderation strategies proposed by *Mistral*, followed closely by its larger version *MistralNemo*, and *Qwen2*. Interestingly, these models were already found to be reasonably aligned in score assignment, justification, and suggestion for improvement during our quantitative evaluation (cf. Section 3.6.1). By contrast, *Llama3* was the least preferred model. Figure 10 illustrates the distribution of model preferences across annotated posts. Notably, for certain questions, questionnaire respondents showed near-consensus agreement toward specific models, suggesting that these models were particularly effective on certain <rules, post> pairs.

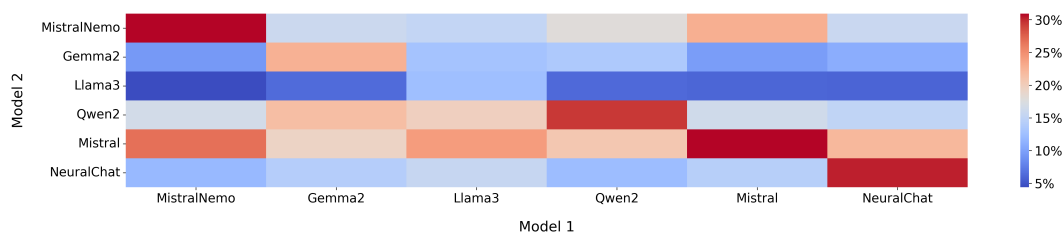

Fig. 11. Agreement matrix between human experts (questionnaire respondents) across the different LLM-moderators. Each cell $(i, j)$ indicates the degree of agreement/overlap between the choices of model $i$ and model $j$. Values on the diagonal $(i, i)$ how many times each model was chosen across all questions. Values outside the diagonal, i.e., cells $(i, j)$ with $i \neq j$, indicate that a fraction of raters who have chosen model $i$ frequently chose also model $j$, i.e., how many times two models were selected together for the same question. Each column sums to one. Warmer, resp. colder, colors indicate higher, resp. lower, percentage values.

In this regard, Figure 11 shows the agreement matrix among models based on human evaluation. While the diagonal confirms our previous findings about the most preferred models, interesting remarks arise from the off-diagonal cells, with some clear patterns emerging. For instance, the questionnaire respondents who preferred *MistralNemo* tend to also prefer its smaller version *Mistral* (and vice versa) confirming the similar moderation capabilities observed throughout this study. Likewise, questionnaire respondents who selected *Qwen2* and *NeuralChat* also prefer *Mistral*, further stressing the alignment in performance among these models.

**Human Perceptions of LLMs' Moderation Strategies.** Here we analyze three key aspects of how humans perceive the moderation strategies based on their chosen LLM for each <rules, post> pair.

Figure 12 (left) shows the Likert-scale distribution of how well the compliance scores assigned by the selected models align with humans' perception of the actual post compliance. Remarkably, most models are strongly in sync with human judgment, with scores consistently above 4 (out of 5) across the entire distribution and similar median values. Interestingly, although less frequently preferred, *Gemma2* and *Llama3* show strong alignment with human perception when selected. As a side remark, the minimum score of 3 out of 5 further strengthens the overall adherence to human expectations.

As reported in Figure 12 (center), the *Mistral* family of models also emerges as the one with the best consistency between the assigned compliance scores and the corresponding justifications. Notably, all models show a minimum consistency rating of 3, with median values exceeding 4, indicating that most models maintain strong coherence between score assignment and the corresponding rationale.

Finally, as illustrated in Figure 12 (right), the *Mistral* family of models are perceived as the most useful, followed by *Qwen2* and *NeuralChat*, consistently with all our previous findings. Interestingly, *Llama3*, although less commonly chosen, achieves recognition of notable usefulness when selected as the preferred model.

**Human Feedback on LLMs' Moderation Strategies.** In the last part of our human-based evaluation, we collected open-ended feedback on the strengths and weaknesses of the preferred LLM agents in performing automatic compliance checking.

Nearly all evaluators praised the effectiveness and accuracy of the moderation, noting that compliance scores were adequately justified, and the suggestions for improving compliance were clear and helpful. They appreciated the models' ability to handle both straightforward and complex posts, their understanding of offensive, harmful, and sensitive content, and their capacity to balance enforcing community rules with maintaining freedom of speech, rather than resorting to censorship.

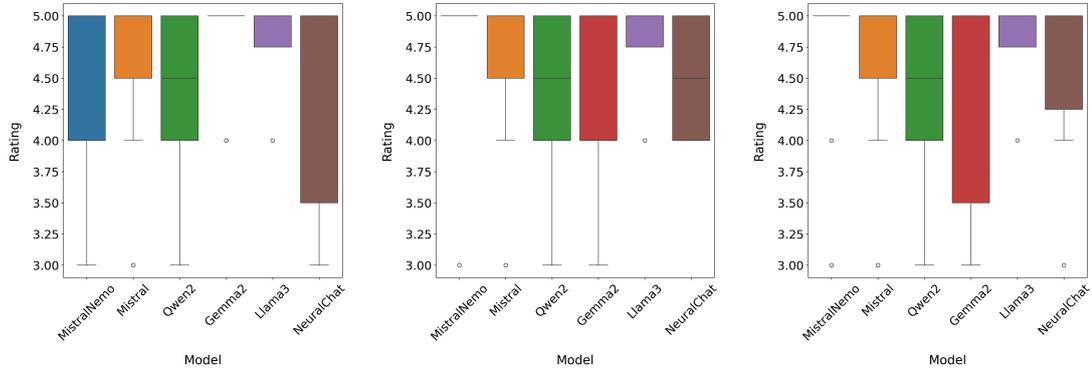

Fig. 12. Distribution of the scores assigned by the domain experts (questionnaire respondents) to the matching between their perception of the compliance level and the one outputted by the preferred LLM-moderator (left), the fitting of the justification text provided by the preferred LLM-moderator with the corresponding score (center), and the usefulness of the preferred LLM-moderator (right).

While the overall feedback was positive, some minor issues were identified. The most common concern was the perceived severity of the scores, with some evaluators finding the moderation too harsh in certain situations or noting occasional instances where scores seemed slightly higher than expected. Additionally, some evaluators suggested improved recognition of "highly sarcastic" content and called for additional content verification measures to prevent the spread of fake news.

## 5 Discussion and Conclusions

The growing volume and rapid spread of social media content have overwhelmed traditional manual moderation approaches, making it increasingly challenging for humans to verify whether posts comply with community guidelines. Earlier attempts to alleviate this burden through automated systems, like Reddit's *Automod*, have struggled to meet human expectations [21]. However, the advent of LLMs and AI agents has opened unprecedented possibilities for tackling this problem.

In this work, we addressed this challenge by evaluating the performance of six Open-LLMs in determining content compliance across more than 50,000 posts from hundreds of Mastodon servers, a previously unexplored yet challenging benchmark in content moderation due to the heterogeneity in community scopes and rules, and the recognized strain in content moderation [1] faced by this paradigm.

Below, we summarize the key findings from our research work, organized around the three research questions we outlined in the Introduction.

**A1: LLMs' Moderation Capabilities.** Both our qualitative and quantitative evaluations demonstrated notable capabilities of the LLM agents in providing effective moderation strategies for communities with publicly available rules. The models exhibit strong flexibility in contextualizing their compliance checking to the communities' rules and scope, tailoring their proposed strategies to diverse scenarios. Also, they provide particularly clear and informative justifications for their assigned scores and offer practical suggestions for improving content compliance with community rules. We found that the selected LLM-moderators are robust to several exogenous factors that could influence their judgments. In this regard, we emphasize that their capabilities are not dependent on rules' verbosity, as they effectively understand the semantics of guidelines even in the presence of very concise rules.

**A2: Consistent Behaviors.** Despite being mostly different in architecture and training data, most LLMs exhibit a certain level of consistency in their automatic compliance checking, as confirmed by the fair-to-moderate inter-rater agreement scores. The models were also found to provide semantically similar score justifications, thus catching similar patterns in content, yet by using different words. Despite less evident, this also holds for the suggestions they provide to improve compliance. Among the six models we evaluated, the *Mistral* family, *Qwen2* and *NeuralChat* emerge as the most consistent and agreeing ones. Notably, these are also the most preferred models by our human evaluators, who tended to choose their strategies interchangeably across different posts, indicating that these models possess very similar moderation-capabilities.

**A3: Human Perception.** The involved domain-experts (questionnaire respondents) found the moderation strategies enforced by the LLM-moderators as being particularly close to their perception of the actual compliance levels. This suggests that LLMs have elevated the quality of automatic moderation to unprecedented levels, thus moving away from previous criticisms [21]. Besides this, the domain experts commended the models for their precision, appropriateness, and adaptability across different contexts. They also stressed the strong coherence between the assigned scores and the corresponding justifications provided by the LLMs, particularly highlighting the informativeness of these justifications. Furthermore, the domain experts perceived the LLMs as highly useful, with average scores consistently above 4 out of 5, indicating a strong potential for their adoption in automatic compliance checking. However, the domain experts also observed that the models sometimes appear to be more severe than expected and noted some limitations in understanding ironic content.

**Limitations.** Like any research, this work comes with its limitations, which we acknowledge to guide further research on the topic. One primary limitation is the scarcity of multi-lingual LLMs available for deployment as agents. Indeed, although English is widely used worldwide, many communities operate in other languages, where our proposed approach can significantly contribute to enhancing community safety. Additionally, in today's social landscape, non-compliant content extends beyond text to include various modalities (e.g., audio, video, images). For example, hateful memes [24] are increasingly used to disseminate harmful and non-compliant content in ways that can circumvent traditional text-based detection methods. Finally, our agents were built using LLMs in their default state, without any fine-tuning or behavioral guidance beyond prompting. While demonstrating their effectiveness "out-of-the-box" is valuable, LLMs are typically trained on human data and may exhibit intrinsic biases that could affect their moderation capabilities, demanding further investigations to address these potential shortcomings.

**Conclusions.** In this work, we introduced a novel automated approach for evaluating social media content compliance with community rules within the complex landscape of Decentralized Social Media. Our study represents the first large-scale evaluation of the compliance-checking capabilities of six Open-LLM agents, setting it apart from previous works that relied on closed and black-box API-based models. Through a combination of quantitative analysis and human-based qualitative evaluations, we demonstrated that these Open-LLM agents can effectively detect non-compliant content while adapting their capabilities to heterogeneous communities. Notably, in contrast with earlier approaches, our proposed method matches human expectations and perceptions of compliance, indicating its potential as a viable solution for future LLM-powered content moderation systems. Additionally, the possibility of deploying these models locally addresses privacy concerns, as no sensitive data needs to be transmitted over the Internet, making them both privacy-compliant and cost-effective. We believe our work can lead to significant advancements in automated content moderation, contributing to a safer social environment.

**Future Work.** Future research will be dedicated to expanding the applicability of our proposed framework to non-English communities and extending our framework to handle multi-modal content, thus being able to detect non-compliant content spreading through other languages or modes. Furthermore, we aim to investigate whether specializing these LLM agents (e.g., through domain-specific training) or aligning them to human preferences (e.g., via Direct Preference Optimization [39]), such as the ones we collected during our qualitative evaluation, can lead to enhanced LLM-moderators.

**Ethics Statement.** • This work explores the potential of LLMs for automatic moderation, aiming to advance the understanding of how these models can help address everyday challenges faced by social media platforms. Our findings demonstrate that LLM agents exhibit strong moderation capabilities, often perceived by humans as being quite "close" to human judgment. However, *we disclaim any responsibility for potential misuse or malicious applications of our research outcomes.* • We recognize that, while social media content is publicly accessible, it may contain personal information, and individuals may not consent to their posts being processed by opaque black-box APIs across the internet. In light of this, we fully respect and endorse these concerns. *To ensure privacy, we anonymized all textual content, removed any Personally Identifiable Information (PII), disregarded multimedia attachments, and processed all data locally using open, locally deployed models. No remote APIs were utilized due to concerns around transparency and control.* • Survey participants were made aware of the potentially sensitive nature of the content they were asked to evaluate and were given the option to opt out before viewing any material. *All data collected was used solely for this research and will not be shared with third parties or repurposed for any other objectives.*

**Transparency and Reproducibility.** To ensure transparency and foster reproducibility, we have fully disclosed all details about the prompts used throughout our work, as well as all specifics for deploying our used models (including model identifiers, frameworks, and hyperparameters). *Furthermore, we are committed to publicly releasing all our developed code for data collection and experiments upon acceptance of this work.*

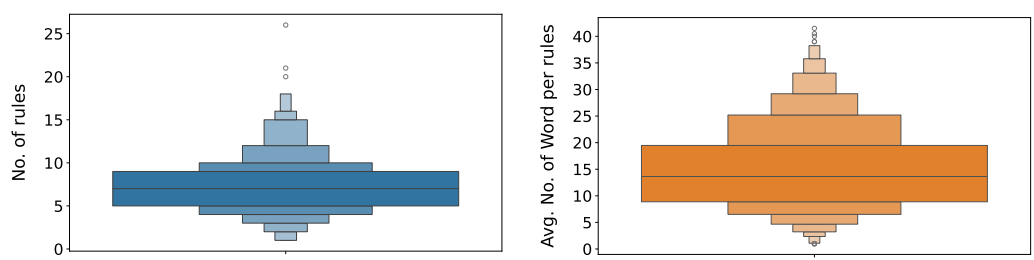

Fig. 13. (On the left) Distribution of the number of rules across our considered Mastodon instances. (On the right) Average number of words used across rule declarations.

## A Additional Details on Server Rules

Mastodon servers vary significantly in both the number and complexity of their rules. As reported in Figure 13 (left), the majority of instances have between 5 and 10 established rules, though there are some outliers with over 25 rules. Additionally, the level of detail in these rules also varies. Figure 13 (right) illustrates that most rule-sets consist of 10 to 15 words, while there are some extreme cases with brief rule-sets containing a couple of words, or lengthy ones containing around 40 words. These servers also vary notably in their context and focus, ranging from broad, general communities to those centered around specific topics. This diversity is reflected in their rule-sets, as shown in Figure 2 (in the main text) and Figure 14 shown below. For this reason, some communities may allow explicit or NSFW content under certain constraints (e.g., content warnings), others may be entirely dedicated to it, while some may strictly forbid it, representing another challenging point for automatic moderation.

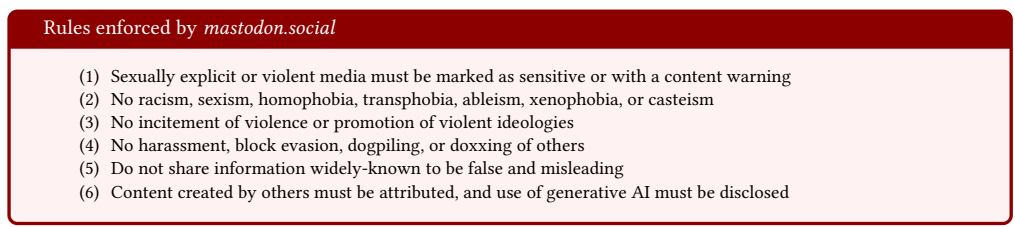

Rules enforced by *mastodon.social*
(1) Sexually explicit or violent media must be marked as sensitive or with a content warning
(2) No racism, sexism, homophobia, transphobia, ableism, xenophobia, or casteism
(3) No incitement of violence or promotion of violent ideologies
(4) No harassment, block evasion, dogpiling, or doxxing of others
(5) Do not share information widely-known to be false and misleading
(6) Content created by others must be attributed, and use of generative AI must be disclosed

Fig. 14. Example of community rules collected from *mastodon.social*.

## B Additional Details on Compliance Scores

In Section 4.1, we observed that most LLMs tend to assign the highest compliance score to a large fraction of posts. We hypothesized that this is due to the large volume of data we collected, which led to the inclusion of older posts (tracing back months), particularly for the smallest instances, which increases the likelihood that non-compliant content has already been moderated. To test this, we analyzed the temporal distribution of posts across compliance score categories for each model. For the sake of brevity, we present the results from *NeuralChat*, the only model also considering the lowest compliance score. As shown in Figure 15, non-compliant posts are concentrated in a shorter and fresher time frame, supporting our hypothesis. This trend also applies to other low compliance scores, typically linked to more recent content.

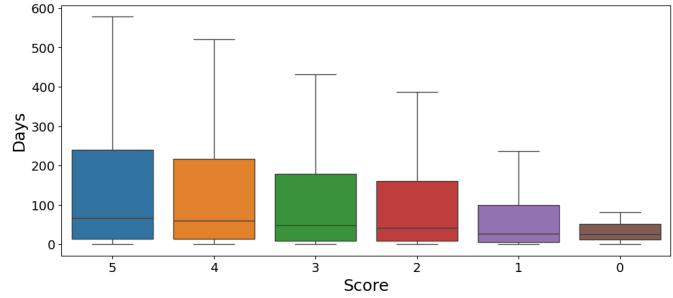

Fig. 15. Temporal distribution of the posts for each score-category as established by NeuralChat. The number of days refers to the delta from the most recent post in our dataset.

Table 5. Binning of the average compliance scores achieved by our data from the six LLM-moderators.

| Score avg. bin | (4.167, 5.0] | (3.333, 4.167] | (2.5, 3.333] | (1.667, 2.5] | (0.829, 1.667] |
|---|---|---|---|---|---|
| No. of posts | 41,844 | 5640 | 1745 | 966 | 539 |

## C Additional Details on Human Evaluation

Table 5 categorizes the average compliance scores expressed by our LLM-moderators across the 50,800 reviewed posts into five distinct bins. From these bins, a random sample of 30 posts (six for each bin) was selected for evaluation by domain experts during the qualitative analysis phase.

Figure 16 also reports the informed consent module provided to each of the domain experts who was involved in the questionnaire, before presenting any content. This module includes all necessary instructions for evaluating the moderation strategies effectively, along with a warning about the possibility of encountering sensitive material.

Finally, Figures 17-18 illustrates the survey's user interface using an example of a <rules, post> pair. Participants were instructed to choose the box containing the moderation approach they perceived as the most appropriate, associated with one of the six LLM moderators we implemented. Additionally, they were asked to provide both quantitative and qualitative feedback on their chosen strategy.

**Mastodon Automatic Moderation - Informed Consent**

Dear Participant,

Thank you for agreeing to take part in our research study!

We are conducting a study on automatic content moderation in Mastodon. The purpose of this study is to understand how users perceive different automatic moderation strategies generated by six Open Large Language Models (LLMs) for Mastodon posts from various instances, based on the adherence to the respective community rules.

This survey is **anonymous** and is expected to take approximately **one hour** to complete. It involves the annotation of **30 posts**.

**Instructions:**
- Each page of the survey will present a set of rules specified by a Mastodon instance. Please read these rules carefully.
- After reviewing the rules, you will be shown a post from that instance.
- You will then see six different moderation strategies suggested by the LLMs (the identity of the models will not be disclosed). You are required to select the best moderation strategy based on the community rules (considering factors like precision, clarity, conciseness, absence of unnecessary details, and relevance of the output).
- Once you have chosen the best strategy, you will evaluate how well its compliance score matches your assessment of the post's adherence to the rules and the quality of the reasoning provided by the model.
- Finally, you will be asked to provide some written feedback (in English) on the strengths and weaknesses of the moderation reasoning. Please note that to minimize bias, you will not be able to change your responses after proceeding to the next page.

**Important Notice:**
This study may contain sensitive content. By continuing, you consent to participate in this study and authorize the use of your annotations for research purposes. The information collected will only be used for this research and will not be shared or used for any other purpose. If you do not wish to participate, please close this window.

Thank you for your participation! Enjoy the study!

Fig. 16. The informed consent module presented to questionnaire respondents before starting to evaluate the automatic moderation strategies.

**Community Rules:**
1. Sexually explicit or violent media must be marked as sensitive when posting 2. No racism, sexism, homophobia, transphobia, xenophobia, ableism, casteism, or any other form of bigotry. 3. No incitement of violence or promotion of violent ideologies 4. No harassment, dogpiling, doxxing or otherwise abusive behavior. 5. No illegal content 6. Do not share intentionally false or misleading information

**Post content:**
Just for the record, I would 100% endorse a slut for President, because I base my choices on their values and the work they've done. You should sleep with as many consenting adults as you want, or none at all, why the fuck would I care?

**Choose the most appropriate moderation approach among the following:**

(Rater #1)
**Score:** 4: Compliant
**Justification for the score:** The post does not violate any of the listed policies.
**Suggestions for improvement:** N/A

(Rater #2)
**Score:** 3: Reasonably Compliant
**Justification for the score:** The post uses offensive language and promotes a potentially harmful view on sexual behavior.

**Suggestions for improvement:** The user should avoid using derogatory language and consider the potential impact of their statements on others.

(Rater #3)
**Score:** 2: Somehow Non-Compliant
**Justification for the score:** The post contains explicit content that is not marked as sensitive, and it promotes sexual freedom in a manner that may be considered inappropriate or offensive to some users.
**Suggestions for improvement:** Mark the content as sensitive and ensure the language used is respectful and inclusive to all users.

(Rater #4)
**Score:** 2: Somehow Non-Compliant
**Justification for the score:** The post contains derogatory language and might be perceived as promoting a certain lifestyle without respecting the diversity of opinions, which violates policy 2.
**Suggestions for improvement:** Rewrite the post to focus on the values and work of the individual without using derogatory language or promoting a specific lifestyle.

(Rater #5)
**Score:** 1: Non-Compliant
**Justification for the score:** The post contains sexist language and promotes a derogatory term for women.
**Suggestions for improvement:** Remove the derogatory term and refrain from using sexist language.

(Rater #6)
**Score:** 1: Non-Compliant
**Justification for the score:** The post contains explicit language and promotes a sexist ideology.

Note: The compliance level is determined based on the policies provided. If a post violates one or more policies, the compliance level is 1. If a post does not violate any policy,
**Suggestions for improvement:** Consider rephrasing the post to avoid explicit language and focus on the values and work of the candidate. ```python
def check_post_compliance(post_text):
    # Define the policies
    policies = {
        1: "Sexually

Fig. 17. Sample screenshot of the user interface of the survey. For each <rules, post> pair, we ask participants to select the most suitable moderation approach from a set of six proposals, corresponding to shuffled and anonymized LLM-moderators.

**Based on your chosen rater, please evaluate the following points:**

|  | 0 | 1 | 2 | 3 | 4 | 5 |
|---|---|---|---|---|---|---|
| Rate how well the given compliance score corresponds to your perception of the compliance of the posts with respect to the rules. | ○ | ○ | ○ | ○ | ○ | ○ |
| Rate how well the justification text provided by the chosen rater fits with the compliance score assigned by the chosen rater. | ○ | ○ | ○ | ○ | ○ | ○ |
| Rate how much useful you evaluate the moderation strategy provided by your chosen rater. | ○ | ○ | ○ | ○ | ○ | ○ |

**Comments/Feedback on the strengths of the chosen moderation strategy:**

**Comments/Feedback on the weaknesses of the chosen moderation strategy:**

Fig. 18. (Cont.) Sample screenshot of the user interface of the survey. Respondents are also asked to rate various aspects of their preferred strategy and provide open-ended textual feedback.